\documentclass[10pt]{article}
\usepackage{amsmath}
\usepackage[dvips]{epsfig}
\usepackage[dvips]{graphics}
\usepackage{a4}
\usepackage{graphicx}
\usepackage{amsfonts}
\usepackage{amssymb}
\begin{document}

\title{Massless DKP fields in Riemann-Cartan space-times}
\author{R. Casana$^{\mathrm{a}}$, V.Ya. Fainberg$^{\mathrm{a}}$ \thanks{ Permanent
address: P. N. Lebedev Institute of Physics, Moscow, Russia }\,\,, J.T.
Lunardi$^{\mathrm{b}}$,\\B.M. Pimentel$^{\mathrm{a}}$, and R.G. Teixeira$^{\mathrm{c}}$\footnote{
Voluntary Professor at Instituto de F\'{\i}sica Te\'orica, Universidade
Estadual Paulista.} \vspace{.2cm}\\{\small $^{\mathrm{a}}$Instituto de F\'{\i}sica Te\'orica, Universidade
Estadual Paulista} \vspace{-.1cm}\\{\small Rua Pamplona 145, CEP 01405-900, S\~ao Paulo, SP, Brazil} \vspace{.1cm}\\{\small $^{\mathrm{b}}$Departamento de Matem\'atica e Estat\'{\i}stica,
Universidade Estadual de Ponta Grossa} \vspace{-.1cm}\\{\small Av. Gal. Carlos Cavalcanti 4748, CEP 84032-900, Ponta Grossa, PR,
Brazil}\vspace{.1cm}\\{\small $^{\mathrm{c}}$Faculdade de Tecnologia e Ci\^encias Exatas,
Universidade S\~ao Judas Tadeu}\vspace{-.1cm}\\{\small Rua Taquari 546, CEP 03166-000, S\~ao Paulo, SP, Brazil}}
\date{}
\maketitle
\begin{abstract}
We study massless Duffin-Kemmer-Petiau (DKP) fields in the context of
Einstein-Cartan gravitation theory, interacting via minimal coupling
procedure. In the case of an identically vanishing torsion (Riemannian
space-times) we show that there exists local gauge symmetries which reproduce
the usual gauge symmetries for the massless scalar and electromagnetic fields.
On the other hand, similarly to what happens with the Maxwell theory, a
non-vanishing torsion, in general, breaks the usual $U(1)$ local gauge
symmetry of the electromagnetic field or, in a different point of view, impose
conditions on the torsion.
\end{abstract}


\section{Introduction}

The \textit{Duffin-Kemmer-Petiau equation} (DKP) is a first order relativistic
wave equation that describes spin 0 and 1 fields \cite{Petiau, Duffin,
Kemmer}. It is analogous to the Dirac equation, but differs by the algebra
satisfied by the $\beta^{\mu}$ matrices, correspondent to the Dirac
$\gamma^{\mu}$ ones, which have only three irreducible representations of
dimensions 1 (trivial), 5 (spin 0) and 10 (spin 1).

In the last years there have been a renewed interest in DKP theory as it was
realized that DKP theory is richer than the KG one with respect to the
introduction of interactions \cite{Kalbermann, Clark}. For instance, it has
been studied in the context of QCD \cite{Gribov}, covariant hamiltonian
dynamics \cite{Kanatchikov}, in the causal approach \cite{Lunardi 3}, in the
context of five-dimensional galilean covariance \cite{Esdras}, in the
scattering $K^{+}$-nucleus \cite{Prog}, etc. On the other hand, there have
been some efforts to give strict proofs of equivalence between the KG
{equation} and the spin 0 sector of DKP {equation} in various situations
\cite{Fainberg 1,Fainberg 2,Fainberg 3}. In the same context, some aspects
regarding the minimal interaction with the electromagnetic field have been
clarified \cite{Lunardi 2, Nowakowski}.

In the context of curved space-times, it was proved the complete equivalence
between DKP and KG and Proca fields in a riemannian space-time \cite{Lunardi
1, Red'kov}. Moreover, it was shown that in the context of Einstein-Cartan
gravity DKP theory naturally induces an interaction between the spin 0 field
and the space-time torsion, which breaks the equivalence with the KG theory
\cite{Lunardi 4}. Besides that, DKP theory induces additional couplings
between spin 1 fields and the space-time torsion, what also breaks the
equivalence with the Proca theory \cite{trabs=1}.

In the present work we develop further the above analysis by studying the
\textit{massless} DKP field minimally coupled to Riemann-Cartan space-times,
as well the gauge invariance properties of the theory. It is important to
notice that the massless case can not be obtained through the limit
$m\rightarrow0$ of the massive case. This is due to the fact that the
projections of DKP field into spin 0 and 1 sectors involve the mass as a
multiplicative factor \cite{Lunardi 4, trabs=1} so that taking the limit
$m\rightarrow0$ makes the results previously obtained useless. Moreover, if we
simply make mass equal to zero in the usual massive DKP Lagrangian we obtain a
Lagrangian with no local gauge symmetry. As will be seen, the solution is to
change the usual mass term in the DKP Lagrangian to a term containing a
singular matrix, what will change the way we manipulate the equation of motion.

So, this paper is organized as follows: in section 2 we present the formalism
introduced by Harish-Chandra for the massless DKP theory in the Minkowski
space-time \cite{HarishC} and select its spin 0 and 1 sectors through the use
of the Umezawa's projectors \cite{Umezawa, Lunardi 2, Lunardi 1}. In section 3
we study massless DKP fields minimally coupled to Riemann-Cartan space-times
and compare the results with those obtained in the context of KG and Maxwell
theories in the presence or not of torsion. In section 4 we present our
remarks and conclusions. The basic aspects and properties of DKP equation
which are necessary to read this work can be found in the references
\cite{Lunardi 1, Lunardi 2}, where it was used the same metric signature
($\eta=diag(+,-,-,-)$) used here.

\section{The massless DKP theory}

As mentioned above, massless DKP theory can not be obtained as a zero mass
limit of the massive DKP case, so we consider the Harish-Chandra Lagrangian
density for the massless DKP theory in the Minkowski space-time $\mathcal{M}%
^{4}$, given by \cite{HarishC}
\begin{equation}
\mathcal{L_{M}}=i\overline{\psi}\gamma\beta^{a}\partial_{a}\psi-i \partial
_{a}\overline{\psi}\beta^{a}\gamma\psi-\overline{\psi} \gamma\psi\;,
\label{eq1}%
\end{equation}
where the $\beta^{a}$ matrices satisfy the usual DKP algebra
\begin{equation}
\beta^{a}\beta^{b}\beta^{c}+\beta^{c}\beta^{b}\beta^{a}=\beta^{a} \eta
^{bc}+\beta^{c}\eta^{ba} \label{eq2x1}%
\end{equation}
and $\gamma$ is a \textit{singular} matrix satisfying\footnote{We choose a
representation in which ${\beta^{0}}^{\dag}={\beta^{0}}$, ${\beta^{i}}^{\dag
}=-{\beta^{i}}$ and $\gamma^{\dag}=\gamma$ \thinspace.}
\begin{equation}
\beta^{a}\gamma+\gamma\beta^{a}=\beta^{a}\quad{\mathrm{and}} \quad\gamma
^{2}=\gamma\,. \label{gama}%
\end{equation}
From the above lagrangian follows the massless DKP wave equation
\begin{equation}
i\beta^{a}\partial_{a}\psi-\gamma\psi=0\;. \label{eq2}%
\end{equation}

The equations (\ref{eq1})-(\ref{eq2}), as it was shown by Harish-Chandra,
describe in fact four massless gauge theories, which correspond to a massless
scalar field, a spin 1 (i.e. electromagnetic) field, a second-rank
antisymmetric potential and a third rank linear potential. The last two ones
propagate no degree of freedom and are topological field theories
\cite{HarishC,reft1,reft2}.

The Lagrangian density (\ref{eq1}) and equation (\ref{eq2}) are both invariant
under the following local gauge transformation
\begin{equation}
\psi\rightarrow\psi^{\prime}=\psi+(1-\gamma)\Phi\,, \label{eq3}%
\end{equation}
where the field $\Phi$ is assumed to satisfy the condition
\begin{equation}
i\,\beta^{a}\partial_{a}(1-\gamma)\Phi=0\,. \label{eqx5}%
\end{equation}
This condition, when projected to the spin 0 and 1 sectors reproduce,
respectively, the global and local U(1) gauge symmetry of the massless
Klein--Gordon and electromagnetic fields, as we will show.

\subsection{Spin $0$ sector \label{s0fin}}

To select the spin 0 sector from equation (\ref{eq2}) we use the Umezawa's
projectors $P$ and $P^{a}$ \cite{Umezawa,Lunardi 2}, remembering that under
proper Lorentz transformations $P\psi$ transforms as a scalar field while
$P^{a}\psi$ transforms as a vector field. Applying these projectors on
equation (\ref{eq2}), and taking into account that relations (\ref{gama})
imply $\gamma P=P\gamma$ and $P^{a}\gamma+\gamma P^{a}=P^{a}$, we obtain the
equation of motion for the massless scalar field $P\psi$
\begin{equation}
\partial_{a}\partial^{a}\left(  P\psi\right)  =0\,. \label{eqq6}%
\end{equation}
In terms of the scalar field $P\psi$ the gauge transformation (\ref{eq3})
reads
\begin{equation}
P\psi^{\prime}=P\psi+(1-\gamma)P\Phi\quad,\quad P^{a}\psi^{\prime} =P^{a}%
\psi+P^{a}(1-\gamma)\Phi\,, \label{eqq9}%
\end{equation}
and condition (\ref{eqx5}) becomes
\begin{equation}
\partial_{a}P^{a}(1-\gamma)\Phi=0\quad,\quad\partial^{a}P(1-\gamma)\Phi=0\,.
\label{eqq10}%
\end{equation}
Therefore, it can be easily verified that equation (\ref{eqq6}) is invariant
under transformation (\ref{eqq9}).

The results above are independent of the representation for the algebra
(\ref{eq2x1})-(\ref{gama}). Nevertheless, to study the gauge invariance in
more details we shall use a specific representation for $\beta^{a}$ in which
\begin{equation}
\gamma=diag(\lambda,1-\lambda,1-\lambda,1-\lambda,1-\lambda).
\end{equation}

In this representation the one-column DKP wave function and its projections
are given by (with $a=0,1,2,3$)
\begin{align}
\psi &  =\left(
\begin{array}
[c]{c}%
\varphi\\
\psi^{a}%
\end{array}
\right)  ,\quad P\psi=\left(
\begin{array}
[c]{c}%
\varphi\\
\left[  0\right]  _{4\times1}%
\end{array}
\right)  ,\quad P\gamma\psi=\left(
\begin{array}
[c]{c}%
\lambda\varphi\\
\left[  0\right]  _{4\times1}%
\end{array}
\right)  ,\label{ap3a}\\
P^{a}\psi &  =\left(
\begin{array}
[c]{c}%
\psi^{a}\\
\left[  0\right]  _{4\times1}%
\end{array}
\right)  ,\quad P^{a}\gamma\psi=\left(
\begin{array}
[c]{c}%
(1-\lambda)\psi^{a}\\
\left[  0\right]  _{4\times1}%
\end{array}
\right)  \label{ap3b}%
\end{align}

The condition $\gamma^{2}-\gamma=0$ will imply for the $\lambda$ parameter
\begin{equation}
\lambda^{2}-\lambda=0\rightarrow\lambda=0,1\,. \label{ap2}%
\end{equation}
The value $\lambda=1$ corresponds to a topological field, while $\lambda=0$
reproduces the massless Klein-Gordon field \cite{HarishC}, as we shall see in
the following. In this representation the explicit relations among the
components of the massless spin $0$ DKP field are
\begin{align}
&  (1-\lambda)\psi^{a}=i\partial^{a}\varphi\,,\nonumber\\
& \label{eqq11}\\
&  \lambda\varphi=i\,\partial_{a}\psi^{a}\,,\nonumber
\end{align}
where $a=0,1,2,3$. If the column matrix $\Phi$ is given by
\begin{equation}
\Phi=\left(  \varphi_{_{\Phi}},\phi^{0},\phi^{1},\phi^{2}, \phi^{3}\right)
^{T} \label{fi}%
\end{equation}
the gauge transformation (\ref{eqq9}) for the components of $\psi$ reads
\begin{align}
\varphi^{\prime}  &  =\varphi+(1-\lambda)\varphi_{_{\Phi}}
\,,\nonumber\label{eqq12}\\
& \\
\psi^{\prime}{}^{a}  &  =\psi^{a}+\lambda\,\phi^{a}\,,\nonumber
\end{align}
while the condition (\ref{eqq10}) gives
\begin{equation}
\lambda\partial_{a}\Phi^{a}=0\quad,\quad(1-\lambda)\partial^{a} \varphi
_{_{\Phi}}=0\,. \label{eqq13}%
\end{equation}
Using the results of the $\lambda=0$ case it can be seen that the DKP
Lagrangian density (\ref{eq1}) reduces to the usual one for the massless
Klein-Gordon field
\begin{equation}
\mathcal{L_{M}}_{_{s=0}}=\partial^{\mu}\varphi^{\ast}\partial_{\mu} \varphi\,,
\label{eqq13x}%
\end{equation}
which, together with the equation of motion $\partial_{a} \partial^{a}%
\varphi=0$, is invariant under the gauge transformation $\varphi^{\prime
}=\varphi+\varphi_{_{\Phi}}$, with constant $\varphi_{_{\Phi}}$.

From equations (\ref{eqq11}) we see that the $\lambda=1$ case corresponds to a
constant field $\varphi$ (thus propagating no degree of freedom). This is a
topological field which will not be considered here. \newline 

\subsection{Spin $1$ sector \label{s1fin}}

To select the spin 1 sector we use the projectors $R^{a}$ and $R^{ab}$, such
that under proper Lorentz transformations $R^{a}\psi$ and $R^{ab}\psi$
transform, respectively, as a vector and a second rank tensor field
\cite{Umezawa,Lunardi 2}. Again, applying these projectors on (\ref{eq2}) and
taking into account that, due to relations (\ref{gama}), we have $\gamma
R^{a}=R^{a}\gamma$ and $R^{ab}\gamma+\gamma R^{ab}=R^{ab}$ we obtain the
equation for the massless vector field $R^{a}\psi$
\begin{equation}
\partial_{b}\left[  \partial^{a}(R^{b}\psi)-\partial^{b}(R^{a}\psi) \right]
=0\,. \label{eqq15}%
\end{equation}
The gauge transformation (\ref{eq3}) and the condition (\ref{eqx5}) yield
\begin{align}
&  R^{a}\psi^{\prime}=R^{a}\psi+R^{a}(1-\gamma)\Phi\quad,\quad R^{ab}%
\psi^{\prime}=R^{ab}\psi+R^{ab}(1-\gamma)\Phi,\label{eqq151}\\
&  \partial_{b}R^{ab}(1-\gamma)\Phi\;=\;0\quad,\quad\partial^{a}R^{b}
(1-\gamma)\Phi-\partial^{b}R^{a}(1-\gamma)\Phi\;=\;0, \label{eqq152}%
\end{align}
and the invariance of (\ref{eqq15}) under the gauge transformation
(\ref{eqq151})-(\ref{eqq152}) can be promptly verified .

Again we shall study the gauge transformation by considering a specific spin 1
representation for $\beta^{a}$ such that
\begin{equation}
\gamma=diag(\lambda,\lambda,\lambda,\lambda,1-\lambda,1-\lambda,
1-\lambda,1-\lambda,1-\lambda,1-\lambda),
\end{equation}
and the condition $\gamma^{2}-\gamma=0$ once more implies that for the
$\lambda$ parameter we have
\begin{equation}
\lambda^{2}-\lambda=0\rightarrow\lambda=0,1\,.
\end{equation}

In this representation the wave function $\psi$ and its projections are given
by
\begin{align}
&  \psi=\left(
\begin{array}
[c]{c}%
\left[  \psi^{a}\right]  _{4x1}\\
\left[  \psi^{ab}\right]  _{6x1}%
\end{array}
\right)  \quad,\quad R^{a}\psi=\left(
\begin{array}
[c]{c}%
\psi^{a}\\
\left[  0\right]  _{9x1}%
\end{array}
\right)  \quad,\quad R^{a}\gamma\psi=\left(
\begin{array}
[c]{c}%
\lambda\psi^{a}\\
\left[  0\right]  _{9x1}%
\end{array}
\right) \label{ap6a}\\
&  R^{ab}\psi=\left(
\begin{array}
[c]{c}%
\psi^{ab}\\
\left[  0\right]  _{9x1}%
\end{array}
\right)  \quad,\quad R^{ab}\gamma\psi=\left(
\begin{array}
[c]{c}%
(1-\lambda)\psi^{ab}\\
\left[  0\right]  _{9x1}%
\end{array}
\right)  \quad,\quad a=0,1,2,3. \label{ap6b}%
\end{align}

As it will be clear below, these choices of $\lambda=0$ correspond to the
Maxwell's equations and $\lambda=1$ to a topological field \cite{HarishC}. The
relations among the $\psi$ field components now are
\begin{align}
&  \lambda\psi^{a}\;=\;i\,\partial_{b}\psi^{ab}\,,\nonumber\label{eqq20}\\
& \\
&  i\,(1-\lambda)\psi^{ab}\;=\;\partial^{a}\psi^{b}-\partial^{b}\psi
^{a}\,.\nonumber
\end{align}
If the matrix $\Phi$ is now written as
\begin{equation}
\Phi=\left(
\begin{array}
[c]{c}%
\left[  \phi^{a}\right]  _{4\times1}\\
\left[  \phi^{ab}\right]  _{6\times1}%
\end{array}
\right)
\end{equation}
the gauge transformation (\ref{eqq151}) becomes
\begin{equation}
\psi^{\prime}{}^{a}=\psi^{a}+(1-\lambda)\phi^{a}\quad,\quad\psi^{\prime}%
{}^{ab}=\psi^{ab}+\lambda\phi^{ab}\,,\label{eqq21}%
\end{equation}
while (\ref{eqq152}) gives
\begin{equation}
\lambda\,\partial_{b}\phi^{ab}=0\quad\,,\quad(1-\lambda)\left(  \partial
^{a}\phi^{b}-\partial^{b}\phi^{a}\right)  =0\,.\label{eqq22}%
\end{equation}
From (\ref{eqq20}) we see that the $\lambda=1$ case corresponds to a
topological field propagating no degree of freedom, which will not be
considered here. On the other hand, the $\lambda=0$ case corresponds to the
Maxwell electromagnetic field, what can be promptly realized from
(\ref{eqq20}) by setting
\begin{equation}
\psi^{a}=\frac{1}{\sqrt{2}}\;A^{a}\,.
\end{equation}
The gauge transformation (\ref{eqq21}-\ref{eqq22}) now reads
\begin{equation}
A^{\prime a}=A^{a}+\phi^{a}\,\quad\,,\quad\phi^{a}=\partial^{a}\Lambda(x)\,,
\end{equation}
where $\Lambda(x)$ is some arbitrary function of space-time coordinates. This
is the usual $U(1)$ gauge invariance.

Finally, taking the explicit form of $\psi$ into the DKP Lagrangian density
(\ref{eq1}) results in the Maxwell's one,
\begin{equation}
\mathcal{L_{M}}_{_{s=1}}=-\frac{1}{4}\;F_{ab}F^{ab},
\end{equation}
\begin{equation}
F_{ab}=\partial_{a}A_{b}-\partial_{b}A_{a}\,.
\end{equation}

%

\section{Transition to Riemann-Cartan space-times}

We can do the transition from the Minkowski space-time, $\mathcal{M}^{4}$, to
the Riemann-Cartan one, $\mathcal{U}^{4}$, through the formalism of
\textit{tetrads} and applying the standard form of the minimal coupling
procedure\footnote{By minimal coupling we mean the change from usual
derivatives to covariant ones as stated in \cite{ref4,hehl,misner}. This is
usually referred as the ``comma to semicolon'' rule.}. From now on the Latin
space-time indexes $a,b,...$ will refer to the Minkowski space-time
$\mathcal{M}^{4}(x)$, which now is tangent to the Riemann-Cartan space-time
$\mathcal{U}^{4}$ at the point $x$, whose coordinates will be labeled by the
Greek letters.

The minimally coupled massless DKP lagrangian becomes
\begin{equation}
\mathcal{L_{U}}=\sqrt{-g}\left[  i\overline{\psi}\gamma\beta^{\mu}{\nabla
}_{\mu}\psi-i{\nabla}_{\mu}\overline{\psi} \beta^{\mu}\gamma\psi
-\overline{\psi}\gamma\psi\right]  \;, \label{eq23}%
\end{equation}
where $g$ is the determinant of the Riemann-Cartan metric tensor $g^{\mu\nu}$
and the matrices $\beta^{\mu}=\beta^{\mu}(x)$ are defined through contraction
with the tetrad (or \textit{vierbein}) fields $e^{\mu}{}_{a}(x)$, i.e.,
$\beta^{\mu}=e^{\mu}{}_{a} \beta^{a}$. These matrices satisfy the generalized
DKP algebra\footnote{We denote by $\eta^{ab}$ the (constant) metric tensor of
$\mathcal{M}^{4}(x)$.}
\begin{align}
&  \beta^{\mu}\beta^{\nu}\beta^{\alpha}+\beta^{\alpha}\beta^{\nu} \beta^{\mu
}=\beta^{\mu}g^{\nu\alpha}+\beta^{\alpha}g^{\nu\mu}\,,\label{eq1b}\\
&  \beta^{\mu}\gamma+\gamma\beta^{\mu}=\beta^{\mu}\quad{\mathrm{and}}
\quad\gamma^{2}=\gamma\,.
\end{align}
In the above lagrangian $\nabla$ is the Einstein-Cartan covariant derivative
associated with a connection $\Gamma_{\mu\,\nu}^{ \quad\alpha}$, whose
antisymmetric part defines the \textit{torsion tensor} $Q_{\mu\,\nu}%
^{\quad\alpha}$, i.e.,
\[
Q_{\mu\,\nu}^{\quad\alpha}=\frac{1}{2}\left(  \Gamma_{\mu\,\nu}^{\quad\alpha}
-\Gamma_{\nu\,\mu}^{\quad\alpha}\right)  \,.
\]
The \textit{contorsion} tensor $K_{\mu\,\nu}^{\quad\alpha}$ is defined as
\[
K_{\mu\,\nu}^{\quad\alpha}=\overset{r}{\Gamma}_{\mu\,\nu}^{ \quad\alpha
}-\Gamma_{\mu\,\nu}^{\quad\alpha}=-Q_{\mu\,\nu}^{ \quad\alpha}+Q_{\mu\;\;\nu
}^{\;\;\alpha}+Q_{\nu\;\;\mu}^{\;\; \alpha}\,,
\]
where $\overset{r}{\Gamma}_{\mu\,\nu}^{\quad\alpha}$ are the Christoffel
symbols, or the Riemannian part of the connection $\Gamma_{\mu\,\nu}%
^{\quad\alpha}$. The covariant derivatives of DKP field are given by
\[
{\nabla}_{\mu}\psi=\partial_{\mu}\psi+\frac{1}{2}\omega_{\mu ab}S^{ab}%
\psi\quad,\quad{\nabla}_{\mu}\overline{\psi}= \partial_{\mu}\psi-\frac{1}%
{2}\omega_{\mu ab}\overline{\psi}S^{ab}\,,
\]
where $S^{ab}=[\beta^{a},\beta^{b}]$ and $\omega_{\mu ab}$ is the spin
connection \cite{Lunardi 4}. In the Einstein-Cartan theory the spin connection
can be written in terms of the affine connection and the tetrad field as
\cite{ref4}
\begin{equation}
\omega_{\mu}{}^{ab}=\gamma_{\mu}{}^{ab}-K_{\mu}{}^{ba}, \label{eq33c}%
\end{equation}
where the term $\gamma_{\mu}{}^{ab}$ in (\ref{eq33c}) is referred to as the
\textit{riemannian part} of the spin connection.\footnote{This term is given
by
\[
\gamma_{\mu}{}^{ab}=-\gamma_{\mu}{}^{ba}=e_{\mu i}\left(  C^{abi}%
-C^{bia}-C^{iab}\right)  \,,
\]
where $C^{abi}$ are the Ricci rotation coefficients
\[
C_{ab}{}^{i}=e^{\mu}{}_{a}\left(  x\right)  e^{\nu}{}_{b}\left(  x\right)
\partial_{\lbrack\mu}{}e_{\nu]}{}^{i}\,.
\]
The brackets in this expression denote symmetrization of the enclosed indexes.}

From the Lagrangian density (\ref{eq23}) we obtain the generalized massless
DKP equation in a Riemann-Cartan space-time
\[
i\,\beta^{\mu}{\nabla}_{\mu}\psi+i\,K_{\mu\nu}^{\quad\!\mu}\beta^{\nu}%
\gamma\psi-\gamma\psi=0\,.
\]
Similarly to what happens with the massive DKP and Dirac fields, the minimal
coupling procedure on the massless lagrangian density leads to a
\textit{non-minimally} coupled equation of motion \cite{ref4,Lunardi 4}. This
equation is invariant under the gauge transformation (\ref{eq3}) if, and only
if, $\Phi$ satisfies
\begin{equation}
\beta^{\mu}{\nabla}_{\mu}(1-\gamma)\Phi=\gamma\beta^{\mu}{\nabla}_{\mu}%
\Phi=0\,,\label{gaugerc}%
\end{equation}
which will be assumed from now on. This condition is nothing more than a
generalization of condition (\ref{eqx5}).

\subsection{Spin $0$ sector}

The Riemann-Cartan spin $0$ projectors $P$ and $P^{\mu}$($=e^{\mu}{}_{a}P^{a}%
$) are such that $P\psi$ and $P^{\mu}\psi$ transform, respectively, as a
scalar and a vector in the Riemann-Cartan space-time \cite{Lunardi 4}.
Following the same procedure done in the Minkowski case we obtain the equation
of motion for the field $P\psi$ in a Riemann-Cartan space-time
\begin{equation}
\left(  {\nabla}_{\mu}+K_{\alpha\,\mu}^{\quad\!\alpha}\right)  ({\nabla}^{\mu
}+K_{\beta}^{\;\;\mu\,\beta}\gamma)(P\psi)=0\,,\label{eq33}%
\end{equation}
as well as the gauge transformation
\begin{equation}
P\psi^{\prime}=P\psi+P(1-\gamma)\Phi\quad,\quad P^{\mu}\psi^{\prime}=P^{\mu
}\psi+P^{\mu}(1-\gamma)\Phi\,.\label{eq34}%
\end{equation}
It is straightforward to verify that equation (\ref{eq33}) is invariant under
the above gauge transformation.

We turn to the specific representation mentioned in the previous section
\ref{s0fin}, and restrict our attention to the $\lambda=0$ case only, which in
the Minkowski space-time represents the usual massless Klein-Gordon field.
Here the relations among the $\psi$ components are
\[
\psi^{\mu}=i{\nabla}^{\mu}\varphi\;=\;i\partial^{\mu}\varphi\,,
\]
such that equation (\ref{eq33}) now reads
\[
({\nabla}_{\mu}+K_{\alpha\,\mu}^{\quad\!\alpha})\,\partial^{\mu}%
\varphi=\overset{r}{\nabla}_{\mu}\partial^{\mu}\,\varphi=0\,.
\]
Thus, contrary to what happens in the massive case \cite{Lunardi 4}, we
conclude that \textit{the spin 0 sector of the massless DKP field does not
interact with the space-time torsion}. With $\Phi$ given by (\ref{fi}) the
gauge transformation (\ref{eq34}) now reads
\[
\varphi^{\prime}=\varphi+\varphi_{_{\Phi}}\quad,\quad\psi^{\prime}{}^{\mu
}=\psi^{\mu}\,,
\]
while condition (\ref{gaugerc}) becomes ${\nabla}_{\mu}\varphi_{_{\Phi}%
}=\partial_{\mu}\varphi_{_{\Phi}}=0$, i.e., $\varphi_{_{\Phi}}$ must be a constant.

Finally, in this representation the lagrangian density for the spin $0$ sector
of the massless DKP field in Riemann-Cartan space-time reduces to the usual
one obtained from the Minkowski Klein-Gordon lagrangian density, i.e.,
\begin{align}
\mathcal{L_{RC}}_{_{s=0}}=\sqrt{-g}\;\partial^{\mu}\varphi^{*} \partial_{\mu
}\varphi\, .
\end{align}

\subsection{Spin $1$ sector}

The Riemann-Cartan spin 1 projectors $R^{\mu}$($=e^{\mu}{}_{a}R^{a}$ ) and
$R^{\mu\nu}$($=e^{\mu}{}_{a}e^{\nu}{}_{b}R^{ab}$) are such that $R^{\mu}\psi$
and $R^{\mu\nu}\psi$ transform respectively as a vector and a second rank
tensor in the Riemann-Cartan space-time \cite{trabs=1}. Following the same
steps of the previous sections we obtain the equation of motion for the field
$R^{\mu}\psi$
\begin{equation}
({\nabla}_{\beta}+K_{\sigma\,\beta}^{\quad\!\sigma})({\nabla}_{\alpha
}+K_{\sigma\,\alpha}^{\quad\!\sigma}\gamma)\,T^{\alpha\beta\mu}\;=\;0\,,
\label{s1p}%
\end{equation}
where $T^{\alpha\beta\mu}=g^{\alpha\beta}(R^{\mu}\psi)-g^{\alpha\mu}
(R^{\beta}\psi)$, and the gauge transformation
\begin{equation}
R^{\mu}\psi=R^{\mu}\psi+R^{\mu}(1-\gamma)\Phi\quad,\quad R^{\mu\nu}
\psi^{\prime}=R^{\mu\nu}\psi+R^{\mu\nu}(1-\gamma)\Phi\,, \label{eq42}%
\end{equation}
together with the conditions
\begin{align}
&  \nabla_{\nu}R^{\mu\nu}(1-\gamma)\Phi=0\,,\nonumber\\
& \label{joao}\\
&  {\nabla}^{\mu}R^{\nu}(1-\gamma)\Phi-{\nabla}^{\nu}R^{\mu}(1-\gamma
)\Phi\;=\;0\,.\nonumber
\end{align}
Again it is straightforward to show that equation (\ref{s1p}) is invariant
under the above gauge transformation.

Working with an explicit representation, as in section \ref{s1fin}, and
setting $\lambda=0$, which in the Minkowski space-time results in the
electromagnetic field, we get the following relations among $\psi$ components
\begin{align}
&  i\,\psi^{\mu\nu}={\nabla}^{\mu}\psi^{\nu}-{\nabla}^{\nu}\psi^{\mu
}\,,\nonumber\\
& \label{eqrcs1l0-1}\\
&  \left(  {\nabla}_{\nu}+K_{\sigma\,\nu}^{\quad\!\sigma}\right)  \psi^{\mu
\nu}=0\,,\nonumber
\end{align}
which leads to the equation of motion for the spin $1$ sector of the massless
DKP field in a Riemann-Cartan space-time
\[
({\nabla}_{\nu}+K_{\sigma\,\nu}^{\quad\!\sigma})\left(  {\nabla}^{\mu}%
\psi^{\nu}-{\nabla}^{\nu}\psi^{\mu}\right)  =0\,.
\]
The gauge transformation (\ref{eq42}) now reads
\[
\psi^{\prime}{}^{\mu}=\psi^{\mu}+\Phi^{\mu}\quad,\quad\psi^{\prime}{}^{\mu\nu
}=\psi^{\mu\nu}%
\]
and the condition (\ref{joao}) becomes
\[
{\nabla}^{\mu}\Phi^{\nu}-{\nabla}^{\nu}\Phi^{\mu}=0\,,
\]
or, explicitly,
\begin{equation}
\partial_{\mu}\Phi_{\nu}-\partial_{\nu}\Phi_{\mu}-2\,Q_{\mu\nu}^{\quad
\!\alpha}\Phi_{\alpha}=0. \label{eqrcs1l0-4}%
\end{equation}
If $Q_{\mu\nu}^{\!\quad\alpha}\equiv0$ (an identically vanishing torsion) the
gauge transformation above reduces to the usual $U(1)$ gauge transformation of
the electromagnetic field, i.e.,
\[
\Phi_{\mu}=\partial_{\mu}\Lambda\,,
\]
where $\Lambda$ is an arbitrary function of the space-time coordinates. If
this is not so, \textit{the condition (\ref{eqrcs1l0-4}), in general, breaks
the usual $U(1)$ local gauge invariance} (by the way, the \textit{global} one
is still preserved). On the other hand, we can find classes of solutions of
(\ref{eqrcs1l0-4}), in terms of torsion, which preserve the local gauge symmetry.

With these results, the spin 1 sector of the massless DKP lagrangian density
in Riemann-Cartan space-time becomes
\[
\mathcal{L_{RC}}_{_{s=1}}=\sqrt{-g}\left(  -\frac{1}{4}\;F_{\mu\nu}F^{\mu\nu
}+F_{\mu\nu}Q_{\quad\!\sigma}^{\mu\nu}A^{\sigma}-Q_{\mu\nu}^{\quad\!\rho
}Q_{\quad\!\sigma}^{\mu\nu}A_{\rho}A^{\sigma}\right)  \,,
\]
where
\[
F_{\mu\nu}=\partial_{\mu}A_{\nu}-\partial_{\nu}A_{\mu}\,,
\]
with $A^{\mu}=\sqrt{2}\psi^{\mu}$ being a real vector field. Clearly, the
terms breaking the $U(1)$ local gauge invariance are those which couple the
torsion to the massless spin 1 field.

%

\section{Remarks and Conclusions}

In this work we considered the massless version of DKP theory obtained from an
explicitly gauge invariant lagrangian density. We generalized the theory from
Minkowski to Riemann-Cartan space-times through the tetrad formalism and
applying the minimal coupling procedure in its standard form, i.e., using the
``comma to semicolon rule". Through the use of Umezawa's projectors we
analyzed the question of the gauge invariance, both in the spin 0 and spin 1
sectors of the theory.

We found that in the spin 0 sector the usual global gauge invariance is
preserved in the transition to Riemann-Cartan space-times (which includes the
Riemannian space-time of general relativity). Moreover, and differently to
what was observed in the massive case \cite{Lunardi 4}, we found that the
massless spin 0 DKP field does not couple with the space-time torsion.
Summarizing, \textit{the spin 0 sector of this massless DKP theory is
completely equivalent to the Klein-Gordon one}.

On the other hand, the spin 1 sector of the theory is invariant under $U(1)$
local gauge transformations, both in the Minkowski and in the Riemannian
space-times, and in these cases it is completely equivalent to the Maxwell
electromagnetic theory. Nevertheless, in Riemann-Cartan space-times with a
nonvanishing torsion there is, in general, a breaking of the local
\textit{$U(1)$} gauge symmetry; although it can be found torsion solutions
which will preserve the local gauge symmetry. These results are similar to
those appearing in the context of Maxwell electromagnetic theory in the
presence of torsion, to which there are several approaches in the literature.
For instance, in reference \cite{hojman} a modified form of the local gauge
invariance is presented, in the context of the minimal coupling procedure,
which allows interaction of the electromagnetic field with a dynamical
torsion. In reference \cite{sabbata} a theory is presented in which the
torsion interacts with the electromagnetic field without modifying the form of
local gauge invariance, but at the cost of introducing a semi-minimal
photon-torsion coupling, justified on the grounds of physical reasonableness.
Alternatively, in reference \cite{Benn}, the authors propose a redefinition of
the minimal coupling rule, such that the gauge invariance is required from the
beginning; and in \cite{hehl2} the minimal coupling procedure is abandoned in
favor of an axiomatic construction based on conservation laws.

As further developments on the massless case, it seems interesting to study
the quantization of spin 1 sector of the massless DKP theory by exploring the
local gauge symmetry. There exist the latent possibility of consistent
quantization in Riemann-Cartan space-time. Besides that, it seems interesting
to study the above subjects in the context of another (nonequivalent) theories
for massless DKP fields, as well as the conformal properties of these fields
on Riemann-Cartan space-times. These questions are presently under our consideration.

%

\subsection*{Acknowledgements}

This work was supported by FAPESP/Brazil (R.C., full support grant 01/12611-7;
V.Ya.F., grant 01/12585-6; B.M.P., grant 02/00222-9), RFFI/Russia (V.Ya.F.,
grant 02-02-16946) and CNPq/Brazil (B.M.P.).

%

\end{document}